\documentclass[9pt,twocolumn,twoside]{pnas-new}

\templatetype{pnasinvited} 

\title{Pandemic Lessons - Devising an assessment framework to analyse policies for sustainability}

\author[a,1]{Pradipta Banerjee}
\author[a,2]{Subhrabrata Choudhury}

\affil[a]{National Institute of Technology Durgapur, India}

\leadauthor{P. Banerjee and S. Choudhury} 

\authorcontributions{}
\authordeclaration{}
\equalauthors{}

\correspondingauthor{\textsuperscript{1}To whom correspondence should be addressed. E-mail: pradiptapb\@email.com}

\keywords{Social Systems $|$ Human Development $|$ Policy $|$ SDG $|$ Ecosystem } 

\begin{abstract}
COVID-19 pandemic has sharply projected the globally persistent multi-dimensional fundamental challenges in securing general socio-economic wellbeing of the society. The problems intensify with increasing population densities and also vary with several socio-economic-geo-cultural activity parameters. 
These problems directly highlight the urgent need for accomplishing the interdependent United Nations Sustainable Development Goals (SDGs) to ensure that in future we do not enter into vicious loops of contracting newer zoonotic viruses and need not search for their vaccines while incurring socio-economic havoc. Behavioural changes in human activities/responses are indispensable for achieving the interdependent SDGs. Using root cause analysis approach, we have developed a yearly assessment framework for viably analysing and identifying requisite region-specific downstream/upstream socio-economic policies to reach the SDGs. The framework makes use of an infographic bar chart representation based on the normalised values of 20 human activity/impact parameters classified under three categories as - negative, limiting and positive. With a holistic view encompassing the SDGs, we illustrate through this framework the impact and urgent need of region-specific human behavioural reforms. This framework enables the foresight about policies regarding their potential in bringing down the negative parameter values to the desired zero level for accomplishing the SDGs through planetary health. \end{abstract}

\dates{}
\doi{\url{}}

\begin{document}

\maketitle
\thispagestyle{firststyle}
\ifthenelse{\boolean{shortarticle}}{\ifthenelse{\boolean{singlecolumn}}{\abscontentformatted}{\abscontent}}{}

\noindent The catastrophic damage caused by COVID-19 has highlighted the underpinnings of several facets of the societal structure and its functioning. This calls for rethinking of policy design frameworks and proactive actions to fight with such adversities. The response to such crisis scenarios maybe categorized into ‘confrontative’ and ‘tactical’ measures. Confrontative ways comprise of medical treatment, provisioning medico-logistics, efforts for curable drugs and vaccines. Surely these measures have to be expedited but they have their own limitations in terms of ratio of established healthcare infrastructure to population, access and distribution of such immobile facilities and minimum time requirement for treatment or for development of vaccines. Hence tactical measures become very crucial to avoid such crisis scenarios so that in future the crises do not crop up in the first place itself. These tactical means can be perceived in terms of both ‘downstream’ and ‘upstream’ measures \cite{1}.

The observations \cite{2,3} that humans have penetrated much deeper into ecozones which were not occupied earlier have got strongly re-emphasized in the context of CoVID-19 pandemic. Such analytics also point to the likelihood of similar outbreaks frequently in future \cite{4} unless requisite societal behavioural changes take place in time. Two major questions that arise in this context are - i) why the sufferings in these crises attain such high intensities and ii) what should be the short to long term measures focussed to address complex social problems so that we do not encounter any pandemic in future be it of contagions, hunger or climate change. This comprehensively relates to the fact that attaining the United Nations Sustainable Development Goals (SDGs) is the only solution to these challenges.

An apparent observation of the SDGs highlights the fact that the goals specifically relate to prevalent societal crisis areas. Policies are designed to achieve the targets associated with these goals. Framing such policies also need to address the issues arising from trade-offs between the SDGs \cite{5}. Evidently this is predominantly a symptomatic approach to resolve the prevalent issues. We propose a root causal approach for assessing and identifying necessary measures to ensure planetary health for achieving the SDGs by investigating the impacts/activities responsible for the prevailing problems. 

The extent of human impacts inflicted on Earth’s geoecological picture are such that even the name of the age we are living in has been based upon human-activities \cite{6}. This is an ominous sign as it relates to the fact that human beings are driving the deadliest mass extinction (also self) by all means \cite{7} and implicitly rendering the existence of a sustainable harmonious social system as impossible. To avert the catastrophes due to such impacts, not only damage control but rather damage-restoration in particular has to be precisely taken up. Villages, farmlands, urbanization, industries and infrastructure development, all have been done at the behest of severe exploitation of nature. The earth system’s balance is already in red and therefore the only possible solution is to rapidly consolidate the present scenario and systematically restore the encroached natural ecozones rather than forcibly creating newer ones. For this purpose, we investigated the basic impacts and driving forces of humans that led to such scenarios.

Probing the causal roots to the existing scenarios we have determined 20 key human activity/impact parameters obtainable from yearly surveys. The info-graphic yearly assessment framework based upon these parameters, having a holistic view encompassing the SDGs, can be used by policymakers and researchers to analyse the ground reality of a specific region and identify the required downstream and upstream measures for sustainability of that region. The rest of this article has been organized as follows: Section 2 (Exploring the roots) provides a comprehensive detailing for root cause analysis of the prevalent interrelated multidimensional crises. The motivation of our approach is greatly influenced by the brighter sides of reduced human activity/response during the pandemic lockdown periods which have been described in Section 3 (Inactivity’s bright picture). Section 4 (Constructing the system monitor) describes the assessment framework primarily focussing on assessing the ‘Why’s instead of right away starting with the ‘What’s and ‘How’s for efficient and sustainable solutions to design the ‘What’s and ‘How’s. The info-graphic bar chart assessment framework along with the correlation of the selected 20 human activity/impact parameters with the SDGs and their breakdown analysis have been illustrated in this section. We conclude our work with a brief discussion about the key concerns for sustainability and how our assessment framework helps in analysing the same in Section 6 (Discussions).

\section*{Exploring the roots}
Apparently exhaustive human actions are continuously destabilizing the ecosystem whose adverse effects are unevenly distributed all across the globe by dint of the simple fact that we all share the same four spheres of the tightly coupled earth system. The instances of such activities present several ironies in the face of human development. Studies \cite{8,9,10} have linked higher COVID-19 morbidity and mortality to air pollution. An interesting fact worth mentioning in this context is that even a short-haul flight produces carbon emission more than a whole year’s carbon footprint of a person from developing nations \cite{11}. Respite from pollution via usage of electric vehicles also seems to be a paradoxical hope apart from questions of its financial viability for masses and battery waste management concerns, because it can contribute even more severely to carbon emission and be hazardous to the environment compared to fossil fuels due to several unanswered critical issues \cite{12}. Solar and nuclear sources also do not provide the perceived green energy alternatives due to their inherent drawbacks and hazards \cite{13,14}. Thereby expedited thrust upon renewable energy production would rather further exacerbate mining threats to biodiversity \cite{15,16}. A reassuring optimistic hope that soon someday a new mass-accessible viable system shall evolve which would produce comparatively nominal polluting levels is hugely detrimental, as the prime question is to rationally think whether humans would survive to ever reach that state given the present rate of nature’s exploitation.

Urban Heat Islands and regional hot-spot formations especially in developing countries pose big challenges globally because of its potential to exacerbate damages inflicted by heatwaves \cite{17,18,19} and precariously disrupt rainfall patterns all across the globe \cite{20,21,22}. Adding to the woes, cooling requirements for air-conditioning, commercial refrigeration and data-centres actually create feedback loops of greenhouse gases as cooling systems themselves are large emitters of these gases thereby temperature rises with increase in usage of cooling systems propelling further cooling requirement. Systems have disbalanced such that as per analysis of almost 40 years’ satellite data \cite{23} the annual snowfall can no longer replenish the melted ice that flows into the ocean from Greenland’s glaciers. 

The adverse socio-economic impacts of natural calamities are also loud and clear \cite{24,25}. In spite of substantial technological advancements, 21\% of world’s food output growth has been lost since 1960s due to climate change \cite{26}. As founded in \cite{27}, one out of every three persons worldwide does not have access to drinking water whereas on the other hand presently the whole world is pushing for stricter hygiene, sanitisation and washing practices. Furthermore, water demand globally is projected to increase by 55\% between 2000 and 2050 \cite{28} while by the year 2040 there will not be enough water in the world to quench the thirst of the world population and keep the current energy and power solutions going if we continue doing what we are doing today \cite{29}. The present agricultural yield is responsible for 70\% of the freshwater usage \cite{30} and food production will need to grow by 69\% by 2035 to feed the growing population \cite{31}. Constraints persist with respect to organic farming \cite{32} which makes it not applicable for several major food crops and has limited potential to decrease water usage. Intriguingly, had the agricultural production been limited only for direct human consumption, the present yield is multiple times more than the existing human population’s total calorie intake requirement \cite{33}. Even then hunger still remains a prevalent crisis in the world due to lack of transparent and efficient distribution mechanisms apart from reasons like that of wars and political crises. In this context studies \cite{34,35} show that not only calorie requirements but also all protein and micronutrient requirements can be easily achieved  without depending on animal sources. Healthy scepticism prevails over in vitro meat as the potential green alternative of conventional meat as it may create more problems than it solves \cite{36,37}. This makes conscious incentivised efforts towards changes in cultivation patterns and food habits \cite{38} crucial as it is evidently unjustified to not evolve accordingly \cite{39,40} amidst such paradoxical crises towards a more humane society barring reasons for mere lack of access. 

Deforestations continue to take place in order to accommodate and create employment opportunities for the growing mass of people. Planting trees in selected lands as per human convenience for compensating the fallouts of urbanisation due to felling of trees from their natural origins raises several pertinent questions. Even after assuming that adequate number of trees have been planted to compensate the amount of deforestation done and they grow ideally upon ensuring of proper maintenance; by the time they become equally efficient carbon sinks the ecoregion demography would sharply change because of the continuously increasing number of carbon emission contributors. Circular Economy models of Make, Use, Reuse, Remake, Recycle, have limited potential in contributing towards sustainability simply due to the restricted scope of reusability and the negative environmental impacts associated with remaking and recycling. The crises may temporarily go down at some place whereas simultaneously increase somewhere else having an uneven distribution across the globe. As evident human societies are running out of time and the coming 50 years would be the decider for future harmonious sustainability. Over the years reactive and symptomatic treatments for socio-ecological problems completely relying on technological innovations have rather resulted in giving rise to newer disorders in place of freeing from the existing problems. This relates to the greater need of policy designs with a global focus having a sustained local region-specific emphasis. All these scenarios imperatively point out to the fact that scientific advancements would not be solely able to redress the prevalent crises and root causal solution approaches have to be figured out to induce calculated, planned yet overhauling societal behavioural changes for eradicating the prevailing crises from its core without any further transgression of the ecozones.

\section*{Inactivity’s bright picture} 
The motivation of our approach is greatly influenced by the brighter sides of reduced human activity/response during the pandemic lockdown periods which have been described

The brighter aspects observed while the whole world was compelled to go into strict lockdown reveals that mother nature had been rapidly recuperating \cite{41} within one-month. The pollution levels went down to that which existed few decades ago while sea and river waters got considerably clean rejuvenating the wildlife. Deeper analysis and rethinking of these phenomena unravels the fundamental truth that infinite development cannot be feasibly pursued on fixed resources. A crucial understanding needs to be perceived that with every unit measure of GDP it accounts for exhaustion of a substantial share of the limited non-renewable natural resources. This purportedly technology backed human development and growth is in actuality continuously destroying the ecological balances propelling climate changes and also making human habitats prone to zoonotic pandemics that of COVID-19. Evidently the invasive nature of human activities is taking its toll on social and ecological welfare. Sharply focussed policies for “planetary health” stands out to be the only probable way-out for human sustenance and egalitarian sustainability.

\section*{Constructing the system monitor}
Taking insights from the discussed crisis scenarios we focussed on assessing the ‘Why’s instead of directly starting with the ‘What’s and ‘How’s for finding efficient and sustainable solutions. For this purpose, we devised an assessment mechanism based on year-wise survey data which would enable to visualize and continuously analyse the overall impact of any policy on the local region-specific demographic structure, environmental status and socio-economic scenario. This region-specific approach has its focus on sustainable global development maintaining planetary health.

To illustrate our root causal analysis approach, we take up the issue of encroachment of ecozones by humans and assess the ‘Why’s associated with it. The underlying reason behind frequent incidents of wild animals intruding human habitat is that in reality humans have invaded their natural homes in the woods. Deforestations have taken place to establish farmlands, villages or for setting up infrastructure and industries by simply destroying the ecosystem balance. Urbanization of regions continuously add on to the imbalances. Apparently in this process we have encroached the natural ecozones beyond permissible limits. All these are done only for providing employment and accommodating increasing human population through development aided by technological advancements and industrializations. Thereby for putting things back to order the only way out is to systematically restore the original ecoregion demographics by focussed reforestations apart from means of damage control through afforestation.

Based upon the root causal analysis approach we have selected 20 human activity/impact parameters crucial for human sustainability in our yearly analysis framework. These parameters based on annual survey data are classified under three categories of - negative, limiting and positive. The selected parameters are given a notation index starting with N, L and P, for being negative, limited and positive respectively, to be used in following tables. The indexed parameters are listed along with their self-explanatory description in Table \ref{First} -  

\begin{table*}
	\centering
	\begin{tabular}{|l|l|l|}
		\hline
		\multicolumn{1}{|c|}{\textbf{Negative}}                                                                                       & \multicolumn{1}{c|}{\textbf{Limiting}}                                                                                         & \multicolumn{1}{c|}{\textbf{Positive}}                                                                                                            \\ \hline
		N1 - Unemployment percentage                                                                                         & L1 – Green cover area growth percentage                                                                              & \begin{tabular}[c]{@{}l@{}} \\P1 - Percentage of population owning \\ personal green cover of 50 sqm per person\end{tabular}                \\
		N2 - Percentage of people undernourished                                                                                  & L2 – Dietary carbon footprint percentage                                                                 & 
		
		\begin{tabular}[c]{@{}l@{}} \\P2 - Percentage of population having habitable \\ surface area of at least 35 sqm per person\end{tabular} \\
		N3 - Percentage of effective population migration                                                                    & \begin{tabular}[c]{@{}l@{}}\\L3 – Carbon footprint percentage from \\ industries and vehicles\end{tabular} &                                                                                                                                          \\ 
		N4 - Incidence of animal abuse cases                                                                                & L4 - Percentage of freshwater usage for Irrigation                                                                         &   
			\begin{tabular}[c]{@{}l@{}} \\ \\ \end{tabular} \\                                                                                              
		\begin{tabular}[c]{@{}l@{}} \\N5 - Incidence of criminal cases of \\ violence against women\end{tabular}                & L5 - Percentage of freshwater usage by Industries                                                                          & 
		\begin{tabular}[c]{@{}l@{}} \\ \\  \end{tabular}                                                                                                                                         \\
		\begin{tabular}[c]{@{}l@{}}\\N6 - Incidence of criminal cases of \\ violence other than against women\end{tabular}     & L6 – Human population growth percentage                                                                    &  
		\begin{tabular}[c]{@{}l@{}} \\ \\ \\ \\ \end{tabular}                                                                                                                                        \\
		N7 - Incidence of Corruption Cases                                                                                   &                                                                                                                       & 
		\begin{tabular}[c]{@{}l@{}} \\ \\ \end{tabular}                                                                                                                                         \\
		N8 - Percentage of illiteracy                                                                                        &                                                                                                                       &          
		\begin{tabular}[c]{@{}l@{}} \\ \\ \end{tabular}                                                                                                                                \\
		\begin{tabular}[c]{@{}l@{}}N9 - Percentage of people not under \\ comprehensive health insurance\end{tabular}        &                                                                                                                       &     
		\begin{tabular}[c]{@{}l@{}} \\ \\ \\ \end{tabular}                                                                                                                                     \\
		\begin{tabular}[c]{@{}l@{}}N10 - Percentage of Hospital Bed occupancy \\ for infectious diseases\end{tabular}        &                                                                                                                       &      
		\begin{tabular}[c]{@{}l@{}} \\ \\ \\ \end{tabular}                                                                                                                                    \\
		N11 - Water Pollution Index (WPI)                                                                                        &                                                                                                                       &               
		\begin{tabular}[c]{@{}l@{}} \\ \\  \end{tabular}                                                                                                                           \\
		\begin{tabular}[c]{@{}l@{}}N12 - Urban Heat Island Index (UHII)\\ (calculated also over rural areas)\end{tabular} &                                                                                                                       &                     
		\begin{tabular}[c]{@{}l@{}} \\ \\ \\ \end{tabular}                                                                                                                     \\ \hline
	\end{tabular}
	\caption{Human activity/impact parameter list alongwith corresponding notation index classified for visualizing region-specific sustainability status through the info-graphic bar chart}
	\label{First}
\end{table*}

\begin{figure*}
		\centering
		\includegraphics[scale=1]{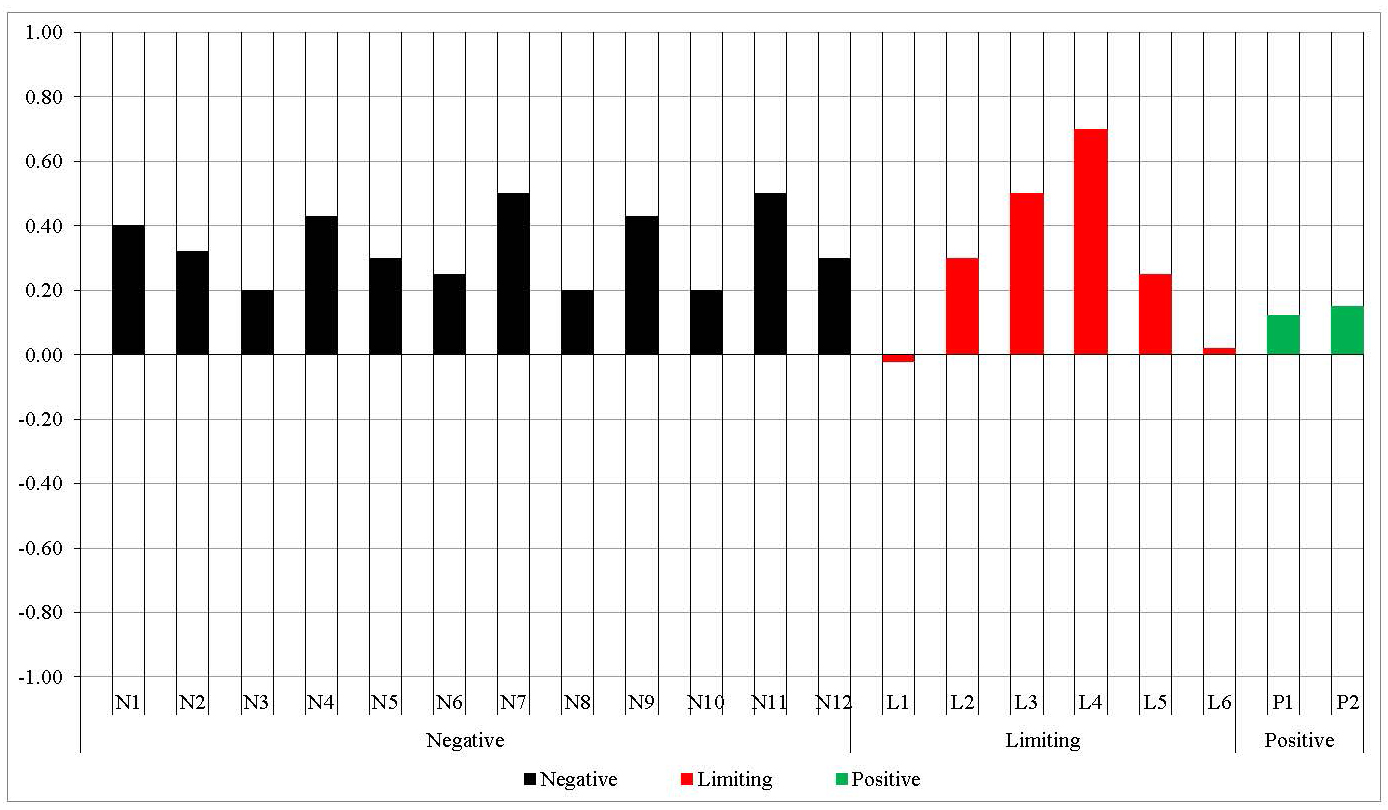}
		\caption{Info-graphic bar chart for assessing region-specific sustainability status}
		\label{A}
\end{figure*}

Incidence of animal abuse, criminal and corruption cases used in N4, N5, N6 and N7 is expressed as the number of events suffered per head of the region’s human population in a year. The two positive parameters introduced in the framework focus on the minimum requirements of a person for her/his mental and physical well-being as emphasized in WHO guidelines \cite{42}. The issues were also enlisted in Agenda 21: Chapter 7: Promoting Sustainable Human Settlement Development \cite{43} however no target had been set for the same then. Lack of organised focus in this dimension has yielded severe blows to the demographic and human settlement structure. Without treating the root causes, planning for living in small congested dwellings for the sake of energy efficiency is a gross violations of the WHO guidelines and the pandemic has reiterated the perils of overcrowding \cite{44}. without  Phenomena like in \cite{45} cannot be just let to happen and systematic policy approaches are required to reverse this scenario globally. We take the parameter for percentage of population both in rural and urban areas owning personal green cover of 50sqm per person by extending the concept of desired Urban Green Space (UGS) of 50sqm per person \cite{46}. Farmlands in rural areas will not be considered for this parameter as basically it is one's workplace. The essential benefits \cite{47,48,49} of personal backyards/gardens would be incorporated through this parameter. Having personal gardens also help in maintaining the required green cover \cite{50} and also signifies the level of decongested habitat one possesses. The second positive parameter for percentage of population having habitable surface area of at least 35sqm per person. Habitable space is considered as the floor area in a dwelling excluding service/utility spaces (kitchen, WC, bathroom, storage, hallways). This has been based upon the Minnnergie A standard of habitable area per person which has been optimised for sustainability through energy efficiency \cite{51}.  Due to heating or cooling requirements in highly developed countries having cold or hot climates respectively the energy consumption per unit area can be considerably high. The value of 35sqm standard as per Minnergie A stands out to be also optimised for such scenarios in highly developed countries like Switzerland even after fulfilling the space criteria for the social and private requirements of an individual for her/his mental and physical welfare as highlighted in WHO guidelines. We have also incorporated two proposed indices of Water Pollution Index (WPI) \cite{52} and Urban Heat Island Index (UHII) \cite{53} as parameters in our assessment framework. The measure of UHII is proposed to be calculated for both urban and rural areas. The info-graphic bar chart utilised to represent these parameters whose values are normalized on a scale of -1 to 1 is demonstrated using Fig. \ref{A} based on the arbitrary illustrative values given in Table \ref{Second}.

\begin{table}
	\centering
	\begin{tabular}{|c|c|}
		\hline
		\textbf{Parameters} & \textbf{\begin{tabular}[c]{@{}c@{}} Values normalised \\ as per the framework's range scale \end{tabular}} \\ \hline
		N1                  & 0.40                                                                                                   \\ \hline
		N2                  & 0.32                                                                                                   \\ \hline
		N3                  & 0.20                                                                                                   \\ \hline
		N4                  & 0.43                                                                                                   \\ \hline
		N5                  & 0.30                                                                                                   \\ \hline
		N6                  & 0.25                                                                                                   \\ \hline
		N7                  & 0.50                                                                                                   \\ \hline
		N8                  & 0.20                                                                                                   \\ \hline
		N9                  & 0.43                                                                                                   \\ \hline
		N10                 & 0.20                                                                                                   \\ \hline
		N11                 & 0.50                                                                                                   \\ \hline
		N12                 & 0.30                                                                                                   \\ \hline
		L1                  & -0.02                                                                                                   \\ \hline
		L2                  & 0.30                                                                                                   \\ \hline
		L3                  & 0.50                                                                                                   \\ \hline
		L4                  & 0.70                                                                                                   \\ \hline
		L5                  & 0.25                                                                                                   \\ \hline
		L6                  & 0.02                                                                                                   \\ \hline
		P1                  & 0.12                                                                                                   \\ \hline
		P2                  & 0.15                                                                                                   \\ \hline
	\end{tabular}
	\caption{Arbitrary illustrative values for the Negative, Limiting and Positive parameters}
	\label{Second}
\end{table}

Values of the negative/positive parameters have a range between 0 to 1. Range of the limiting parameters have been set as -1 to 1 because L1 (Green cover area growth percentage) and L6 (Human population growth percentage) can bear negative values. The limiting parameters represent the core human impacts/activities which needs to be intervened by policy makers through downstream/upstream measures with the objective of maximising the positive parameters while simultaneously bringing down the negative ones to the desired zero level for accomplishing the SDGs through planetary health. This framework enables the foresight in assessing cumulative outcomes of the policies while focusing towards this target to ensure that – i) the limiting parameters, whose values may have an interim allowance for fluctuations under controlled downstream interventions, should be in decreasing order on a medium to long term basis and ii) the positive parameters should always have an increasing trend without ever decreasing. There exists direct correlation between the negative/positive parameters and the interdependent SDGs on a fundamental level, which has been illustrated in Table \ref{Third}. For readability reasons the directly related Negative/Positive Parameters are represented using their corresponding notation index given in Table \ref{First}. From descriptions of the interrelated parameters in Table \ref{First} we infer the direct influencers for each of them to form Table \ref{Fourth}. As because unemployment, school leaving \cite{54} and population size \cite{55} have direct impacts on  N5 (Incidence of criminal cases of violence against women) and N6 (Incidence of criminal cases of violence other than against women)  N1(Unemployment percentage), N8 (Percentage of illiteracy) and L6 (Human population growth percentage) are listed as their direct influencers. N8 (Percentage of illiteracy) is marked as the direct influencer for N9 (Percentage of people not under
comprehensive health insurance) due to the fact that education brings about the awareness for health insurance necessities.

\begin{table*}
	\begin{minipage}{.55\textwidth}
		\centering
		\small
		\begin{tabular}{|l|l|}
			\hline
			\textbf{Sustainable Development Goal} & \begin{tabular}[c]{@{}l@{}} \textbf{Parameters}\end{tabular} \\ \hline
			1. No Poverty                              & N1                     \\ \hline
			2. Zero Hunger                             & N1, N2                     \\ \hline
			3. Good Health and Well-being              & N1, N2, N9, N10, P1, P2    \\ \hline
			4. Quality Education                       & N1, N2, N8                 \\ \hline
			5. Gender  Equality                        & N5, N8                 \\ \hline
			6. Clean Water and Sanitation              & N1, N11                \\ \hline
			7. Affordable and Clean Energy             & N1, N2                 \\ \hline
			8. Good Jobs and Economic Growth           & N1, N8                 \\ \hline
			9. Industry, Innovation and Infrastructure & N1, N8                 \\ \hline
			10. Reduced Inequalities                   & N1, N8                 \\ \hline
			11  Sustainable Cities and Communities     & N2, N3, N4, N5, N6, N7, P1, P2 \\ \hline
			12. Responsible Consumption and Production & N11, N12                \\ \hline
			13. Climate Action                         & N3, N11, N12           \\ \hline
			14. Life Below Water                       & N4, N11                \\ \hline
			15.  Life on Land                          & N2, N4    \\ \hline
			16. Peace, Justice and Strong Institutions & N4, N5, N6, N7         \\ \hline
		\end{tabular}
		\caption{The directly related positive/negative parameters of the assessment framework corresponding to each of the Sustainable Development Goals}
		\label{Third}        
	\end{minipage}\qquad
	\begin{minipage}{.55\textwidth}
		\centering
		\small    
		\begin{tabular}{|l|l|}
			\hline
			\textbf{Parameters} & \textbf{Direct Influencers} \\ \hline
			N1                                  & L6                                                            \\ \hline
			N2                                  & N1, L6                                                 \\ \hline
			N3                                  & N1, L6                                                        \\ \hline
			N4                                  & L2, L6                                                        \\ \hline
			N5                                  & N1, N8, L6                                                    \\ \hline
			N6                                  & N1, N8, L6                                                    \\ \hline
			N7                                  & N1, N8, L6                                                        \\ \hline
			N8                                  & N1                                                            \\ \hline
			N9                                  & N1, N8                                                        \\ \hline
			N10                                 & N1, N2, L1                                                        \\ \hline
			N11                                 & L4, L5, L6                                                    \\ \hline
			N12                                 & L1, L2, L3, L6                                                    \\ \hline
			P1                                  & L6                                                            \\ \hline
			P2                                  & L6                                                            \\ \hline
		\end{tabular}
	    \hspace{5cm}
		\caption{Mapping of direct influencers for \\ each of the Negative/Positive parameters}
		\label{Fourth}        
	\end{minipage}
\end{table*}

From this table an elemental level view of the core human activities/impacts can be drawn from the parameters directly influencing the sustainable development goals. Upon breakdown analysis of the direct influencers we find out that the 6 limiting human activities/impacts form the core control of the negative and positive parameters which in turn directly govern the SDGs. Further among these 6 limiting parameters L6 (Human population growth percentage) emerge as the most widely influential ones.

Analysing the issue of GHGs it is evident that not only reducing per capita footprint is crucial but the cumulative output is even more concerning simply because the positive impacts of reduction in per capita emission through technological innovations can be reversed by the increasing number of GHG producers. Every human being is essentially a gigantic sized consumer in comparison of all other species, consuming and also permanently exhausting considerable amount of the Earth’s limited natural resources. Population growth in the underdeveloped countries is much higher than in the developed ones \cite{56} and as projected in \cite{57} 97\% of the global population growth of 1.2 billion people will be in the developing world between 2013 and 2030. With increasing average life expectancy the issue needs more systematic focus. The nearly vertical growth of the world population from around 1700 AD \cite{58}, after the Industrial Revolution picked up,  which radically changed the global habitat demography by exploiting the environment surely has much deeper socio-economic constructs other than just technological and healthcare advancements. It is also evident that post Industrial Revolution, avenues and growth of income from developing human capital through personal skills overtook that from private wealth like inherited ownership of land. The transformation processes of rags to riches and vice-versa became eminent. Thereby personal aspirations, growing competition with future insecurities, gender discriminative socio-economic customs, lack of guaranteed social security benefits and weakening of cohesive community social structure towards nuclear family based belief systems can be logically ascertained as the reasons behind the steep rise in population size.

\section*{Discussion}

The vital need of the hour is to engender non-congested rural and urban human habitats where people would not be living in pigeonholed houses bereft of minimum required personal open spaces and gardens indispensable for one’s mental/physical wellbeing while simultaneously returning lands deforested for cities and villages to the woods in a gradual systematic pattern. However even if the population growth and unemployment rates decrease absolute values of their sizes keep on increasing. Considering the periods before Covid-19 pandemic struck, it is evident from the growing global unemployed population size that the present eligible workforce for both production and service sector is many times more than it should ideally have been given the present technological scenario even without accounting for future innovations. With passing of time this ratio is worsening in several dimensions aggravating parameters from inequalities to environmental damages. Clearly it is understandable that with even present technology in hand there remains no scope for the argument that there would be labour shortage if somehow the population pyramid base shrinks sharply. Countries where the absolute number of births are on the rise, they not only are deteriorating their own demographics but are also having adverse impacts on other countries through multiple dimensions from migrations for employment to propelling climate changes as understandable from the illustrations about our shared Earth system in the previous sections. It is a fact that in recent past the absolute number of births have somewhat declined in some of the non-developed countries. However, these can be mostly accredited to reasons arising from the scenarios of economic stress for sustenance more than a conscious effort and evidently by the time the nominal declining trend had started those countries are already overcrowded with the damage have being done. Clearly only populations having much smaller and stable sizes would enable settling for a stable pricing and income equilibrium preserving planetary sustainability. 

Understandably no further mode of denialsism \cite{59} and laissez-faire attitude towards the most influential limiting parameters apparent from our root causal analysis assessment framework is acceptable if the planet has to be saved for egalitarian human sustainability. The growth of total population size of the countries which are not highly developed needs to have a sharp declining curve similar to how steeply it rose post-industrial revolution. Gradual decline dynamics after the natural development of those countries in due course of time would not serve the purpose. Concerted efforts for behavioural changes, ranging from dietary habits to family planning, collectively amounting to cultural reforms are essential to evolve as a more humane civilization. Imperatively without any delay we need focussed activism to build movements for spreading awareness breaking outmoded thinking \cite{60} to create a better world. These movements have to be complimented by policies having a balance of incentives and deterrents which would provision greater social securities for people to allay socio-psychological fears about future uncertainties. Critical emphasis is required to be given upon moral and civic education more than that over knowledge for skill development and physical sciences. Bringing about radical reforms for property inheritance rights would also play a vital role in strengthening the cohesive communityhood framework to bring about the paradigm shift from “My family is my morld” towards “The world is my family”. As per our assessment framework when values of all the negative parameters become zero, the socio-ecological balance for planetary health would be achieved eradicating the problems of scarcity, confrontations and crimes therein by instilling a feeling of stability, security and affluence amongst all individuals of the society. This would also result in quantitatively attaining the ten central capabilities towards human development \cite{61} as per the capability theory of justice by Martha Nussbaum which is arguably the most systematic, extensive, and influential philosophical model for human development till date.




\end{document}